\documentclass[twocolumn,showpacs,preprintnumbers,amsmath,amssymb]{revtex4}
\usepackage{graphicx}
\usepackage{epsfig}
\usepackage{dcolumn}
\usepackage{bm}
\begin{document}
\preprint{}
\title{Shrinking Magnetic Vortices in V$_3$Si Due to Delocalized
Quasiparticle 
Core States: Confirmation of the Microscopic Theory for Interacting Vortices}
\author{J.E.~Sonier$^{1,2}$, F.D.~Callaghan$^{1}$, R.I.~Miller$^{3}$,
E.~Boaknin$^{4}$,
L.~Taillefer$^{2,5}$, R.F.~Kiefl$^{2,6}$, J.H.~Brewer$^{2,6}$, K.F.~Poon$^{1}$
and J.D.~Brewer$^{1}$}
\affiliation{$^1$Department of Physics, Simon Fraser University, Burnaby,
British Columbia V5A 1S6, Canada \\
$^2$Canadian Institute for Advanced Research, Toronto, Ontario, Canada \\
$^3$Department of Physics \& Astronomy, University of Pennsylvania,
Philadelphia, PA 19104 \\
$^4$Department of Applied Physics, Yale University, New Haven, CT
06520-8284 \\
$^5$D\'{e}partment de Physique, Universit\'{e} de Sherbrooke, Qu\'{e}bec
J1K 2R1, Canada \\
$^6$Department of Physics and Astronomy, University of British Columbia,
Vancouver, British Columbia V6T 1Z1, Canada}
\date{\today}
\begin{abstract}
We report muon spin rotation measurements on the conventional type-II
superconductor 
V$_3$Si that provide clear evidence for changes to the inner structure of 
a vortex due to 
the delocalization of bound quasiparticle core states. The experimental
findings 
described here confirm a key prediction of recent microscopic theories
describing 
interacting vortices. The effects of vortex-vortex interactions on the
magnetic and 
electronic structure of the vortex state are of crucial importance to the 
interpretation 
of experiments on both conventional and exotic superconductors in an
applied magnetic field.
\end{abstract}
\maketitle
In 1964, a breakthrough paper by Caroli, de Gennes and Matricon \cite{Caroli} 
showed that in the framework of the microscopic theory, quasiparticles
(QPs) bound to an 
{\em isolated} vortex of a conventional $s$-wave type-II superconductor 
occupy discrete energy levels. Twenty-five years later, 
localized vortex core states were observed for the first time in NbSe$_2$ 
by scanning tunneling microscopy (STM) \cite{Hess}.
Our understanding of the vortex state in type-II superconductors has
accordingly 
progressed from Abrikosov's initial prediction \cite{Abrikosov} based on
the macroscopic 
Ginzburg-Landau (GL) theory \cite{Ginzburg}, to current theories
describing the 
electronic structure of magnetic vortices on a microscopic level. However,
it is only in recent years that predictions have emerged from the microscopic 
theory on the effects of {\em vortex-vortex interactions}. 
In analogy with bringing atoms close together to form a conducting solid, 
increasing the vortex density by applying a stronger magnetic field $H$
enhances the overlap of bound state wave functions of neighboring vortices, 
resulting in the formation of energy bands that allow the intervortex
transfer of QPs 
\cite{Pottinger,Tesanovic,Ichioka:99a,Ichioka:99b}. This is expected to
strongly influence 
experiments on conventional superconductors that are sensitive to
quasiparticle 
excitations, such as specific heat and thermal transport, and to have
profound 
effects on the magnetic structure of the vortex state. However,
understanding the potential interplay between vortices and quasiparticles 
is also
of crucial importance in the study of high-temperature superconductors.
In these and other exotic superconductors, comparatively little is known 
about the structure of the vortex state and its effect on experiments in large
magnetic fields. It is therefore essential to have a solid 
understanding of the behavior of interacting vortices in conventional
superconductors, 
and to establish the connections with quasiparticle properties.

The effect of delocalized QP core states on the spatial
variation of the pair potential $\Delta(r)$ at a vortex site has been 
considered in the framework of the quasiclassical Eilenberger 
theory \cite{Ichioka:99a,Ichioka:99b,Golubov}. These calculations 
show that the effect of the intervortex transfer of QPs on $\Delta(r)$ 
leads to a reduction of the size of the vortex cores with increasing $H$. 
Such shrinking of the vortex cores has in fact been observed by muon spin 
rotation 
($\mu$SR)
\cite{Sonier:97a,Sonier:97b,Sonier:99a,Sonier:99b,Kadono,Ohishi,Price}
and STM \cite{Hartmann}, and proposed as a possible explanation for the
anomalous low-field 
behavior observed of the specific heat in conventional 
superconductors \cite{Sonier:99c,Nohara}. However, experimentally 
there has been inadequate evidence in support of a causal relationship 
between the size of the vortex cores and the delocalization of bound QP
core states.

An important finding has come by way of recent low-temperature 
thermal conductivity measurements, which are sensitive only to extended 
or delocalized electronic excitations. These studies have 
revealed the existence of highly delocalized QPs down to low magnetic 
fields in the vortex state of LuNi$_2$B$_2$C \cite{Boaknin:01},
YBa$_2$Cu$_3$O$_{6.9}$ 
\cite{Chiao}, NbSe$_2$ \cite{Boaknin:03} and MgB$_2$ \cite{Sologubenko}.
In the extreme gap anisotropy superconductors LuNi$_2$B$_2$C and 
YBa$_2$Cu$_3$O$_{6.9}$, the dominant contribution to the thermal conductivity 
is believed to be a field-induced Doppler shift of the QP spectrum
outside the 
vortex cores \cite{Volovik}. Most surprising are the results for NbSe$_2$, 
long believed to be representative of a simple conventional $s$-wave 
superconductor. The high degree of QP delocalization in NbSe$_2$ appears 
to arise from two-gap superconductivity \cite{Boaknin:03}, 
as is the case for MgB$_2$ \cite{Sologubenko}. In these superconductors,
the heat at low 
fields is carried by the QP excitations associated with the smaller of
the two 
energy gaps. This smaller gap is a possible source of the very large
low-field value of 
the core size observed in NbSe$_2$ by $\mu$SR \cite{Sonier:97a} and in 
MgB$_2$ by STM \cite{Eskildsen}. On the other hand, the large core size
observed at low fields in borocarbide \cite{Ohishi,Price} and cuprate 
\cite{Sonier:97b,Sonier:99a,Sonier:99b} superconductors is not completely 
understood.

The conventional superconductors V$_3$Si and Nb provide a unique
opportunity to 
experimentally observe the effects of intervortex QP transfer on the inner 
structure of a vortex. In contrast to the abovementioned superconductors, 
thermal conductivity 
measurements \cite{Boaknin:03,Lowell} reveal no appreciable
delocalization of QP 
states at low $H$ --- as expected when the vortices are 
nearly isolated. V$_3$Si is particularly suitable, since it has a simple 
cubic crystal structure, a relatively high superconducting transition 
temperature ($T_c \! = \! 17$~K) and a large upper critical field 
($H_{c2} \! \approx \! 185$~kOe). To determine the size of 
the vortex cores we measured the internal magnetic field distribution in
V$_3$Si 
by $\mu$SR at the Tri-University Meson Facility (TRIUMF), Vancouver, Canada.
The experiment was performed by implanting spin-polarized muons, which
stop randomly 
in the sample on the length scale ($\sim 10^2$-10$^3$~\AA) of the vortex
lattice (VL). 
The magnetic moment of the muon precesses about 
the local magnetic field $B$ with a Larmor frequency given by $\nu \! =
\! \gamma_\mu B$, 
where $\gamma_\mu \!  = \! 0.0852$~$\mu$s$^{-1}$G$^{-1}$ is the
gyromagnetic ratio 
of the muon. The spatial distribution of $B$ is determined by measuring
the time evolution 
of the muon spin polarization via the 
anisotropic distribution of decay positrons \cite{Sonier:00}.

Small-angle neutron scattering \cite{Yethiraj} and recent STM
\cite{Sosolik} images of 
V$_3$Si show that for a field applied along the fourfold [001] axis, the VL 
undergoes a transition from hexagonal to square symmetry. Kogan {\it et
al.} \cite{Kogan} 
have developed a phenomenological London model that attributes this
transition to 
nonlocality of the relation between the supercurrent density {\bf j}({\bf r}) 
and the vector potential {\bf A} in a region around the vortex core. 
This establishes a connection between the vortex structure and the anisotropy 
of the Fermi surface, which in V$_3$Si produces fourfold symmetry near
the vortex cores. 
We note that a more general model has recently been developed for fourfold 
symmetric superconductors, which incorporates the effects of both Fermi
surface 
and gap anisotropy on the vortex structure \cite{Nakai}. At fields below
7.5~kOe where 
the intervortex spacing is large, the isotropic magnetic 
repulsion of vortices yields a hexagonal lattice. 
With increasing field there is an increased overlap of the fourfold
symmetric regions of the 
individual vortices, and the VL evolves into a square at 40~kOe. 

The Kogan model uses an arbitrary cutoff factor to overcome 
the logarithmic divergence of the internal magnetic field $B(r)$ at the
vortex sites. 
We find that the $\mu$SR data are fit much better using a GL analog of
the Kogan model, 
which properly accounts for the finite size of the vortex cores
\cite{Yaouanc}. 
The spatial profile of the periodic local magnetic field 
in the GL formalism is given by
\begin{equation}
B(r) = B_0 (1-b^4) \sum_{ {\bf G}}
\frac{e^{-i {\bf G} \cdot {\bf r}} \, \, u \, K_1(u)}{\lambda^2 G^2 + 
\lambda^4(n_{xxyy} \, G^4 + d \, G_x^2 G_y^2)} \, .
\label{eq:field}
\end{equation}
Here $b \! = \! B/B_{c2}$ is the reduced field, $B_0$ is the average
internal field, 
{\bf G} are the reciprocal lattice vectors, 
$K_1(u)$ is a modified Bessel function, 
$u^2 \! = \! 2 \xi^2 G^2 (1 + b^2)[1-2b(1 - b)^2]$,  $\xi$ is the GL
coherence length, 
and $n_{xxyy}$ and $d$ are dimensionless parameters arising from the
nonlocal corrections. 
The quartic $n_{xxyy}$ term is an isotropic correction, whereas the
biquadratic $d$ term 
controls the fourfold anisotropy.

\begin{figure}
\centerline{\epsfxsize=3.4in\epsfbox{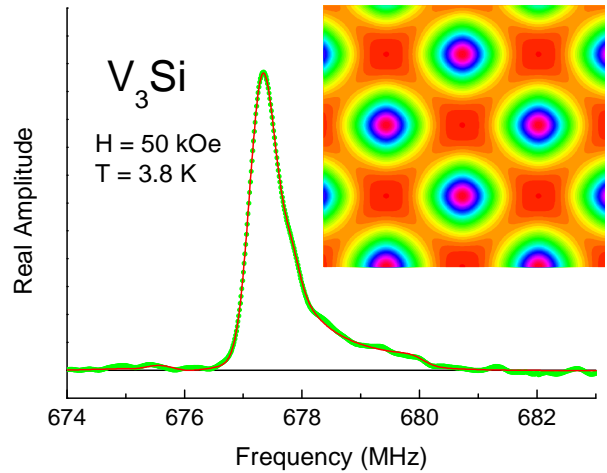}}
\caption{ Fourier transform (FT) of the measured muon spin precession signal (green circles) in V$_3$Si at $T \! = \! 3.8$~K and $H \! = \! 50$~kOe. The FT provides an approximate illustration of the internal magnetic field distribution, but is broadened by the apodization procedure used to smooth out the ringing effects of the finite time window (0-3~$\mu$s) and the noise from reduced counts at later times. Additional broadening due to VL disorder and nuclear magnetic dipoles is accounted for by a Gaussian broadening width of $\sigma \! \approx \! 1.0$~$\mu$s$^{-1}$. The solid red curve is the FT of the fit in the time domain, assuming the field profile $B(r)$ given by Eq.~(\ref{eq:field}). The inset is a contour plot of the function $B(r)$ obtained from the fit.}
\end{figure}

Figure~1 shows Fourier transforms of both the measured muon-spin precession
signal and the fit to the data in the time domain at 50~kOe applied
parallel to [001]. 
In agreement with the imaging
experiments, we obtained excellent fits to a square and hexagonal VL
above 40~kOe
and below 7.5~kOe, respectively, and to a rhombic unit cell with an apex
angle $\beta \! \approx \! 60^{\circ}-90^{\circ}$ at intermediate fields.
These fits and those to Kogan's model also yielded the following results:
(i) For all $H$, $n_{xxyy} \! \approx \! 0$. This is consistent with an
analogous
model developed by Affleck, Franz and Amin \cite{Affleck}, who found that 
the quartic
correction makes a negligible contribution. (ii) Below 7.5~kOe, excellent 
fits were obtained
with $d \! = \! 0$. This indicates that the VL is not strongly tied to the
underlying crystal lattice, as was determined by STM. (iii) Above 7.5~kOe,
the orientation of the vortex cores determined by our analysis
(see Fig.~1 inset) is that which is expected for close packing of square
vortices.

\begin{figure}
\centerline{\epsfxsize=4.0in\epsfbox{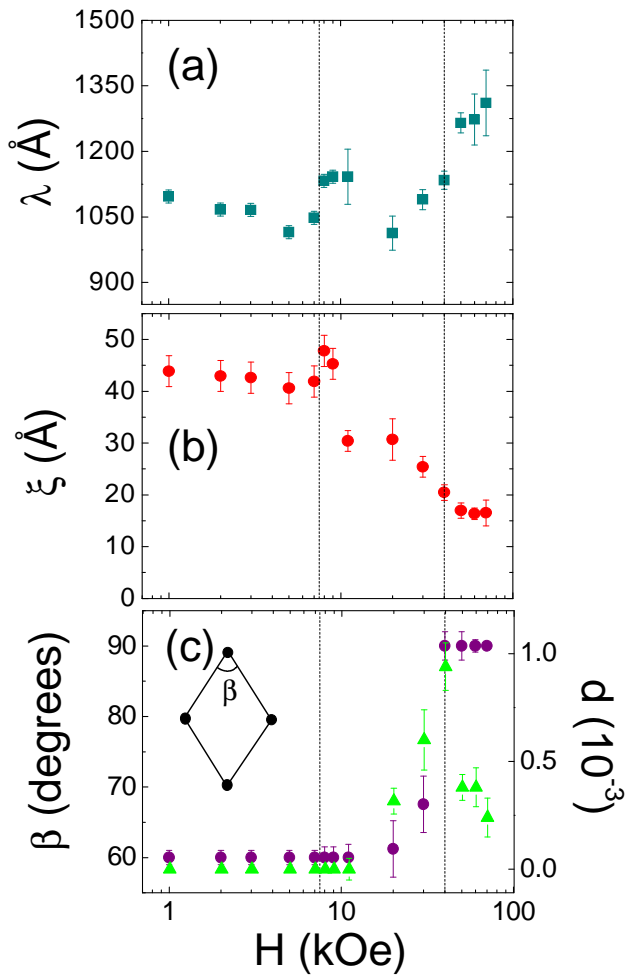}}
\caption{ The magnetic field dependence of the fit parameters at $T \! = \!$ 3.8 K. (a) The magnetic penetration depth $\lambda$. (b) The coherence length $\xi$. (c) The apex angle $\beta$ (purple circles) and the anisotropy parameter $d$ (green triangles). The dashed vertical lines indicate the field range over which the VL continuously evolves from hexagonal to square symmetry. At fields immediately above 7.5~kOe, $d$ and $\beta$ are poorly determined, because the VL
is only slightly distorted from hexagonal symmetry.}
\end{figure}

The field dependence of  $\lambda$, $\xi$, $d$ and $\beta$ are shown in
Fig.~2.
At low fields where the lattice is hexagonal,
$\lambda \! \approx \! 1060$~\AA~and $\xi \! \approx \! 42$~\AA,
which are consistent with previously determined values of these parameters.
Above 7.5~kOe there is a slight nonphysical increase in the fitted values of
$\lambda$ and $\xi$, reflecting the beginning of 
the gradual transition to a square VL.
At higher fields $\xi$ decreases, and continues to do so when the VL
completes its transition to a square at 40~kOe.

Gygi and Schl\"{u}ter \cite{Gygi} have shown that the pair potential
$\Delta(r)$, which in general is a numerical function, varies on two
length scales, 
$\xi_1$ and $\xi_2$. The first is defined from
the slope of $\Delta(r)$ near the center of the vortex core
\begin{equation}
\xi_1 = \Delta_0 / \lim_{r\to0} \frac{\Delta(r)}{r}
\label{eq:core}
\end{equation}
($\Delta_0$ is the BCS superconducting energy gap), whereas $\xi_2$
is the length scale over which $\Delta(r)$ rises to its asymptotic value
$\Delta_0$.
It is important to note that while at low temperatures $\xi_2$ is close
to the value of
the coherence length derived from $H_{c2}$ ({\it i.e.} $\sim 42$~\AA), $\xi_1$
is considerably shorter [see Fig.~3(b)]. Thus, we can understand the
reduction of 
$\xi$ above 7.5~kOe as a rapid variation of $\Delta(r)$ at the vortex center,
and not an increase of $H_{c2}$. Theoretically, the rapid increase of
$\Delta(r)$ near 
$r \! = \! 0$ is accompanied by a similar rise of the supercurrent
density $j(r)$ in the same region \cite{Ichioka:99a,Ichioka:99b,Gygi}.
Consequently, we can define an effective vortex-core size $r_0$ as the
distance
from the core centre ($r \! = \! 0$) to the location where $j(r)$ reaches 
its maximum
value, measured along the line connecting nearest-neighbor vortices.
Solutions of the microscopic theory show that $\xi_1$ and $r_0$ exhibit
the same
qualitative behavior as functions of $H$ and $T$.
From previous $\mu$SR studies of the Kramer-Pesch effect, it is known
that the 
fitted value of $\xi$ in Eq.~(\ref{eq:field}) and $r_0$ exhibit similar
behavior, and 
hence $\xi$ reflects a particular sensitivity to $\xi_1$.  
The parameter $r_0$ can be obatined in a nearly model-independent way
using the Maxwell relation,
$j(r) \! = \! | {\bf \nabla} \times {\bf B}({\bf r})|$, where $B(r)$ is
obtained from fitting
the $\mu$SR time spectrum. Because $r_0$ is not a fit parameter, the
details of the theoretical 
model for $B(r)$ are not very important in this procedure.
What is essential is that the model for $B(r)$ yields excellent fits, as was 
the case using Eq.~(\ref{eq:field}).
\begin{figure}
\centerline{\epsfxsize=3.4in\epsfbox{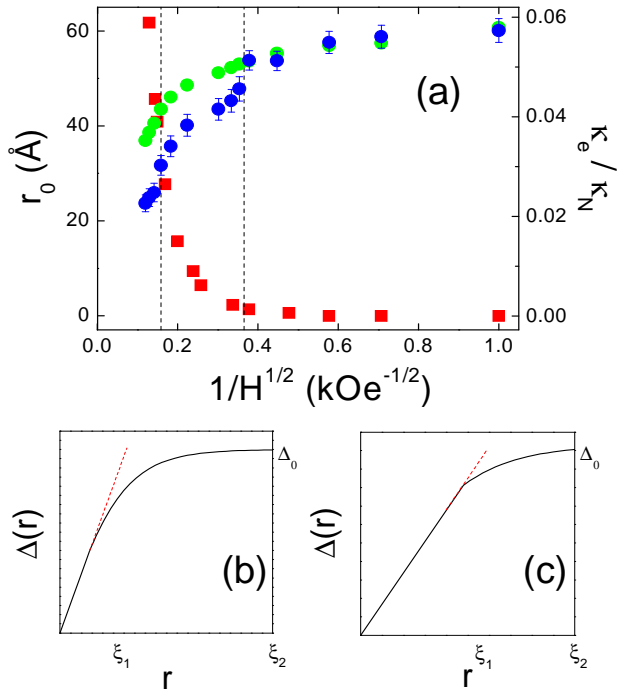}}
\caption{(a) The magnetic field dependence of the vortex core size $r_0$ measured 
by $\mu$SR (blue circles) at $T \! = \! 3.8$~K, and the electronic thermal 
conductivity $\kappa_e/T$ (red squares) from Ref.~\cite{Boaknin:03} 
extrapolated to $T  \! \rightarrow \! 0$ (and normalized to the value $\kappa_N/T$ at $H_{c2}$).
The green circles indicate the reduction of $r_0$ due to the superposition 
of $j(r)$ profiles from individual vortices, calculated assuming the
low-field value of $\xi$.
The data are plotted as a function of $1/H^{1/2}$, which is proportional to the intervortex
spacing $L \! = \! (\Phi_0 / H \sin \beta)^{1/2}$, where $\Phi_0$ is the magnetic flux quantum.
The dashed vertical lines indicate the field range over which the VL undergoes
a continuous hexagonal-square transition. Schematic of the pair potential $\Delta(r)$
as a function of distance from the vortex core centre at high (b) and low (c) magnetic
field.}
\end{figure}

In Fig. 3(a) we compare the field dependence of $r_0$ in V$_3$Si to the
electronic
thermal conductivity $\kappa_e$ measured previously \cite{Boaknin:03}.
For the first time we see the expected field dependence of the core size in a
single-gap conventional $s$-wave superconductor. In contrast to NbSe$_2$
\cite{Sonier:97a} 
and MgB$_2$ \cite{Eskildsen}, the low-field value of $r_0$ is consistent
with the
coherence length calculated from $H_{c2}$. At low fields, where only a modest
overlap of the QP core states of neighboring vortices is expected,
both $r_0$ and $\kappa_e$ exhibit a weak dependence on $H$. The change in 
$r_0$
over this range of $H$ is primarily due to the superposition of the $j(r)$
profiles of nearest-neighbor vortices. This is indicated by the green
circles in Fig. 3. 
Above 7.5~kOe the core size shrinks more rapidly than that expected from
the superposition of $j(r)$ profiles, and is accompanied 
by a simultaneous increase in the electronic thermal conductivity.
Together these observations
signify a change in the slope of $\Delta(r)$
at the vortex center, due to an increased overlap of the bound state wave 
functions
of adjacent vortices. As shown in Fig. 3(a), the transformation to a
square vortex
lattice begins at the field where the delocalization of QPs becomes
significant.
In other words, it is the increased strength of the vortex-vortex
interactions that
drives the symmetry change of the lattice to reflect the fourfold
symmetry of the
individual vortices.

The experimental results presented here for V$_3$Si confirm one of the key
predictions of theoretical works that advocate the importance of intervortex
QP transfer to the detailed structure of the VL. This important detail 
should be a consideration in the interpretation of any experiment on a 
type-II superconductor in an applied magnetic field.

We thank Tetsuo Fukase for providing us with the sample of V$_3$Si.
This work was supported by the Natural Sciences and Engineering Research
Council (NSERC) of Canada and the Canadian Institute for Advanced
Research (CIAR).


\end{document}